\documentstyle[12pt]{article}
\def\be {\begin{equation}}
\def\ee {\end{equation}}
\def\ba {\begin{eqnarray}}
\def\ea {\end{eqnarray}}

\def\bi {\begin{itemize}}
\def\ei {\end{itemize}}
\begin{document}
\def\bea{\begin{eqnarray}}
\def\eea{\end{eqnarray}}
\title{\bf {Corrections to the Cardy-Verlinde formula from the generalized uncertainty
principle }}
 \author{M.R. Setare  \footnote{E-mail: rezakord@ipm.ir}
  \\{Physics Dept. Inst. for Studies in Theo. Physics and
Mathematics(IPM)}\\
{P. O. Box 19395-5531, Tehran, IRAN }}

%\date{\small{\today}}
\maketitle
\begin{abstract}
In this letter, we compute the corrections to the Cardy-Verlinde
formula of $d-$dimensional Schwarzschild black hole. These
corrections stem from the generalized uncertainty principle. Then
we show, one can taking into account the generalized uncertainty
principle corrections of the Cardy-Verlinde entropy formula by
just redefining the Virasoro operator $L_0$ and the central
charge $c$.
 \end{abstract}
% \begin{document}
\newpage
% \vspace*{10mm}

 \section{Introduction}
 It is commonly believed that any valid theory of quantum gravity
 must necessary incorporate the Bekenestein-Hawking definition of
 black hole entropy \cite{bek,haw} into its conceptual framework.
 However, the microscopic origin of this entropy remains an enigma
 for two reasons. First of all although the various counting
 methods have pointed to the expected semi-classical result, there
 is still a lack of recognition as to what degrees of freedom are
 truly being counted. This ambiguity can be attributed to most of
 these methods being based on dualities with simpler theories,
 thus obscuring the physical interpretation from the perspective of
 the black hole in question. Secondly, the vast and varied number
 of successful counting techniques only serve to cloud up an
 already fuzzy picture.\\
  The Cardy-Verlinde formula proposed
   by Verlinde \cite{Verl}, relates the entropy of a  certain CFT with its total
energy and its Casimir energy in arbitrary dimensions. Using the
AdS$_{d}$/CFT$_{d-1}$ \cite{AdS} and dS$_{d}$/CFT$_{d-1}$
correspondences \cite{AS} , this formula has been shown to hold
exactly for different black
holes (see for example {\cite{odi}-\cite{set2}}).\\
Black hole thermodynamic quantities depend on the Hawking
temperature via the usual thermodynamic relations. The Hawking
temperature undergoes corrections from many sources:the quantum
corrections \cite{maj1x}, the self-gravitational
corrections\cite{hawking1}, and the
corrections due to the generalized uncertainty principle.\\
The generalized uncertainty principle corrections are not tied
down to any specific model of quantum gravity; these corrections
can be derived using arguments from string theory \cite{amati} as
well as other approaches to quantum gravity \cite{magi}.\\
In this letter we concentrate on the corrections due to the
generalized uncertainty principle. In section $2$ we review the
connection between uncertainty principle and thermodynamic
quantities, then we drive the corrections to these quantities due
to the generalized uncertainty principle \cite{gupdas}. In
section $3$ we consider the generalized Cardy-Verlinde formula of
a $d-$dimensional Schwarzschild black hole\cite{klem,yum}, then we
obtain the generalized uncertainty principle corrections to this
entropy formula.

\section{The generalized uncertainty principle}
A $d-$dimensional Schwarzschild black hole of mass $M$ is
described by the metric \be ds^{2}=-(1-\frac{16\pi G_d
M}{(d-2)\Omega_{d-2}c^{2}r^{d-3}})c^{2}dt^{2}+(1-\frac{16\pi G_d
M}{(d-2)\Omega_{d-2}c^{2}r^{d-3}})^{-1}dr^{2}+r^{2}d\Omega_{d-2}^{2},
\label{metr} \ee where $\Omega_{d-2}$ is the metric of the unit
$S^{d-2}$ and $G_{d}$ is the $d-$dimensional Newton's constant.
Since the Hawking radiation is a quantum process, the emitted
quanta must satisfy the Heisenberg uncertainty principle \be
\Delta x_{i}\Delta p_{j}\geq \hbar \delta_{ij}, \label{prin} \ee
where $x_{i}$ and $p_{j}$, $i,j=1...d-1$, are the spatial
coordinates and momenta, respectively. By modelling a black hole
as a $d-$dimensional cube of size equal to twice its Schwarzschild
radius $r_{s}$, the uncertainty in the position of a Hawking
particle at the emission is \be \Delta x\approx
2r_{s}=2\lambda_{d}(\frac{G_d M}{c^{2}})^{1/(d-3)}, \label{delta}
\ee where \be
\lambda_{d}=(\frac{16\pi}{(d-2)\Omega_{d-2}})^{1/(d-3)}. \label
{lambda} \ee Using Eq.(\ref{prin}), the uncertainty in the energy
of the emitted particle is \be \Delta E\approx c\Delta
p\approx\frac{M_{pl}c^{2}}{2\lambda_{d}}
m^{-1/(d-3)},\label{dele}\ee where $m=\frac{M}{M_{pl}}$ is the
mass in Planck units and $M_{pl}=(\frac{\hbar
^{d-3}}{c^{d-5}G_d})^{1/(d-2)}$ is the $d-$dimensional Planck
mass. $\Delta E$ can be identified with the characteristic
temperature of the black hole emission, i.e. the Hawking
temperature. Setting the constant of proportionality to
$(d-3)/2\pi$ we get \be T=\frac{(d-3)}{4\pi
\lambda_{d}}M_{pl}c^{2}m^{-1/(d-3)}. \label{temp1} \ee The
entropy is  \be S=\frac{4\pi
\lambda_{d}}{d-2}m^{(d-2)/(d-3)}=\frac{(d-3)}{(d-2)}\frac{Mc^{2}}{T}.\label{entro}
\ee We now determine the corrections to the above results due to
the generalized uncertainty principle. The general form of the
generalized uncertainty principle is \be \Delta x_{i}\geq
\frac{\hbar}{\Delta p_{i}}+\alpha^{2}l_{lp}^{2} \frac{\Delta p_{i}
}{\hbar}, \label{genral}\ee where $l_{pl}=(\frac{\hbar
G_d}{c^{3}})^{1/(d-2)}$ is the Planck length and $\alpha$ is a
dimensionless constant of order one. There are many derivations
of the generalized uncertainty principle, some heuristic and some
more rigorous. Eq.(\ref{genral}) can be derived in the context of
string theory \cite{amati}, non-commutative quantum mechanics
\cite{magi}, and from minimum length consideration \cite{gary}.
The exact value of $\alpha$ depends on the specific model. The
second term in r.h.s of Eq.(\ref{genral}) becomes effective when
momentum and length scales are of the order of Planck mass and of
the Planck length, respectively. This limit is usually called
quantum regime. Inverting Eq.(\ref{genral}), we obtain \be
\frac{\Delta x_{i}
}{2\alpha^{2}l_{pl}^{2}}[1-\sqrt{1-\frac{4\alpha^{2}l_{pl}^{2}}{\Delta
x_{i}^{2} }}]\leq \frac{\Delta p_{i}}{\hbar}\leq \frac{\Delta
x_{i}
}{2\alpha^{2}l_{pl}^{2}}[1+\sqrt{1-\frac{4\alpha^{2}l_{pl}^{2}}{\Delta
x_{i}^{2} }}] \label{momengen} \ee The corrections to the black
hole thermodynamic quantities can be calculated by repeating the
above argument. Setting $\Delta x=2r_{s}$ the generalized
uncertainty principle-corrected Hawking temperature is \be
T'=\frac{(d-3)\lambda_{d}}{2\pi
\alpha^{2}}m^{1/(d-3)}[1-\sqrt{1-\frac{\alpha^{2}}{\lambda_{d}^{2}m^{2/(d-3)}}}]M_{pl}c^{2}
\label{tempcorr}  \ee Eq.(\ref{tempcorr}) may be Taylor expanded
around $\alpha=0$: \be T'= \frac{(d-3)}{4\pi
\lambda_{d}}m^{-1/(d-3)}[1+\frac{\alpha^{2}}{4\lambda_{d}^{2}m^{2/(d-3)}+...}]M_{pl}c^{2}
\label{expan} \ee
\section{Generalized Uncertainty Principle Corrections to the Cardy-Verlinde Formula}
The entropy of a $(1+1)-$dimensional CFT is given by the
well-known Cardy formula \cite{Cardy} \be
S=2\pi\sqrt{\frac{c}{6}(L_0-\frac{c}{24})}, \label{car} \ee where
$L_0$ represent the product $ER$ of the energy and radius, and
the shift of $\frac{c}{24}$ is caused by the Casimir effect. After
making the appropriate identifications for $L_0$ and $c$, the
same Cardy formula is also valid for CFT in arbitrary spacetime
dimensions $(d-1)$ in the form \cite{Verl} \be S_{CFT}=\frac{2\pi
R}{d-2}\sqrt{E_c(2E-E_c)}, \label{cardy}
 \ee the so called Cardy-Verlinde formula, where $R$ is the radius of the system,
 $E$ is the total energy and $E_c$ is the Casimir
 energy, defined as
 \be E_c=(d-1)E-(d-2)TS.  \label{casi} \ee So far, mostly asymptotically AdS and dS
 black hole solutions have been considered \cite{AdS}-\cite{set2}. In \cite{klem},
 it is shown that even the Schwarzschild and Kerr black hole solutions, which are
 asymptotically flat, satisfy the modification of the Cardy-Verlinde formula
 \be S_{CFT}=\frac{2\pi R}{d-2}\sqrt{2EE_c}. \label{cardy1} \ee This result holds also
 for various charged black hole solution with asymptotically flat spacetime \cite{yum}\\
 In this
section we compute the generalized uncertainty principle
corrections to the entropy of a $d-$dimensional Schwarzschild
black hole described by the Cardy-Verlinde formula
Eq.(\ref{cardy1}). The Casimir energy Eq.(\ref{casi}) now will be
 modified due to the uncertainty principle corrections as
\be E'_{c}=(d-1)E'-(d-2)T'S'.  \label{casi1} \ee It is easily
seen that \be
2E'E'_{c}=2(d-1)E'^{2}-2(d-2)E'T'S'=\frac{8(d-1)\pi^{2}
T'^{2}}{(d-3)^{2}}-\frac{4\pi(d-2)T'^{2}S'}{d-3}.
\label{correc}\ee We substitute expressions (\ref{casi1})and
(\ref{correc}) which were computed to first order in $\alpha^{2}$
in the Cardy-Verlinde formula in order that generalized
uncertainty principle corrections to be considered, \bea
\label{cvcor} S'_{CFT}&=&S_{CFT}[1+\frac{\pi
T}{(d-3)E_{c}E}(\frac{4\pi \Delta T}{d-3}-(d-2)T \Delta
S-2(d-2)\Delta TS) ]  \eea where \be \Delta
T=\frac{(d-3)\alpha^{2}m^{\frac{-3}{d-3}}}{16\pi
\lambda_{d}^{3}}M_{pl}c^{2} \label{tempcor}  \ee and \be \Delta
S=\frac{-\pi \alpha^{2}m^{\frac{d-4}{d-3}}}{(d-4)\lambda_{d}},
\hspace{1cm} d>4 \label{entcor} \ee We would like to express the
modified Cardy-Verlinde entropy formula in terms of the energy and
Casimir energy, therefore rewrite the $T, S, \Delta T, \Delta S$
in terms of energy as following \be T=\frac{(d-3)E}{2\pi} \label
{tem1} \ee \be S=\frac{2\pi
(d-3)}{(d-2)}(2\lambda_{d})^{3-d}(M_{pl}c^{2})^{d-2}E^{2-d},
\label{ent1} \ee \be \Delta T=\frac{(d-3)}{2\pi(M_{pl}c^{2})^{2}
}\alpha^{2}E^{3}, \label{temcor1} \ee \be \Delta S= \frac{-\pi
\alpha^{2}}{(d-4)\lambda_{d}}(\frac{2\lambda_{d}E}{M_{pl}c^{2}})^{4-d}.
\label{entcor1} \ee To obtain the last equation we have used the
Eqs.(\ref{dele},\ref{entcor}). Then, Eq.(\ref{cvcor}) can be
rewritten as \bea
\label{cvcor1}S'_{CFT}&=&S_{CFT}[1+\frac{\alpha^{2}}{2E_c}[\frac{2E^{3}}{(M_{pl}c^{2})^{2}}
+\\ &&
(d-3)E^{5-d}(M_{pl}c^{2})^{d-4}(2\lambda_{d})^{3-d}(\frac{d-2}{d-4}-2(d-3))]]
\nonumber \eea As we saw in above discussion these corrections are
caused by generalized uncertainty principle.\\
For the schwarzschild black holes, the dual CFT lives on a flat
space, and thus the energy has no subextensive part. Since the
Casimir energy vanishes, the Cardy-Verlinde formula(\ref{cardy})
makes no sense in this case. In the two-dimensional conformal
field theory, when the conformal weight of the ground state is
zero, we have \be S=2\pi \sqrt{\frac{cL_0}{6}}, \label{car2} \ee
If we use $E_c R=\frac{(d-2)S_c}{2\pi}$ in (\ref{cardy}), where
$S_c$ is the Casimir entropy, and drop the subtraction of $E_c$
in analogy with Eq.(\ref{car2}), we obtain the generalization to
$(d-1)$ dimensions, \be  S=\frac{2\pi}{d-2} \sqrt{\frac{cL_0}{6}}
\label{car3} \ee where $L_0=ER$ and
$\frac{c}{6}=\frac{(d-2)S_c}{\pi}=2E_cR$. Then, we can taking
into account the generalized uncertainty principle corrections of
the Cardy-Verlinde entropy formula by just redefining the
Virasoro operator and the central charge as following \be
L'_{0}=E'R=\frac{R\lambda_{d}}{
\alpha^{2}}m^{1/(d-3)}[1-\sqrt{1-\frac{\alpha^{2}}{\lambda_{d}^{2}m^{2/(d-3)}}}]M_{pl}c^{2}
\label{LEQ} \ee \bea c'&=&12E'_{c}R=\frac{12R\lambda_{d}}{
\alpha^{2}}m^{1/(d-3)}([1-\sqrt{1-\frac{\alpha^{2}}{\lambda_{d}^{2}m^{2/(d-3)}}}]M_{pl}c^{2}\\
&& [(d-1)-(d-2)(d-3)(\frac{\alpha}{\lambda_{d}})^{d-2}I(1,d-4,
\frac{\lambda_{d}m^{1/d-3}}{\alpha})]) \nonumber \label{LEQ1} \eea
Also the first order corrections to the $L_0$ and $c$ are given by
\be \Delta
L_0=L'_{0}-L_0=(E'-E)R=\frac{R\alpha^{2}}{8{\lambda_{d}}^{3}m^{3/(d-3)}}M_{pl}c^{2}=
\frac{\alpha^{2}}{4{\lambda_{d}}^{2}m^{2/(d-3)}}L_0 \label{l1eq}
\ee \bea\Delta
c&=&c'-c=12R(E'_{c}-E_{c})=12R\\
&&[\frac{(d-1)\alpha^{2}}{\lambda_{d}^{3}m^{3/(d-3)}}
M_{pl}c^{2}+(2\lambda_{d})^{3-d}(M_{pl}c^{2})^{d-4}\alpha^{2}E^{5-d}(\frac{d-2}{d-4}-(d-3))]
\nonumber \eea Thus, this redefinition can be considered as a
renormalization of the quantities entering in the Cardy formula.
  \vspace{3mm}

\section{Conclusion}
In this paper we have examined the effects of the generalized
uncertainty principle in the generalized Cardy-Verlinde formula.
The general form of the generalized uncertainty principle is
given by Eq.(\ref{genral}). Black hole thermodynamic quantities
depend on the Hawking temperature via the usual thermodynamic
relation. The Hawking temperature undergoes corrections from the
generalized uncertainty principle as Eq.(\ref{tempcorr}). Then we
have obtained the corrections to the entropy of a dual conformal
field theory live on flat space as Eqs.(\ref{cvcor},\ref{cvcor1}).
Then we have considered this point that the Cardy-Verlinde
(generalized Cardy-Verlinde) formula is the outcome of a striking
resemblance between the thermodynamics of CFTs with
asymptotically Ads (flat) dual's and CFTs in two dimensions.
After that we have obtained the corrections to the quantities
entering the Cardy-Verlinde formula:Virasoro operator and the
central charge. The corresponding problem for the
Schwarzschild-AdS metric is in progress by the author.

\section*{Acknowledgement }
I would like to thank Prof. Saurya Das for  reading the
manuscript also for useful discussions and suggestions.

\end{document}